\documentclass[12pt]{article}
\usepackage{amsmath}
\usepackage{amsfonts}
\usepackage{mathrsfs}
\usepackage{amssymb}
\def\D{\Delta}

\def\dis{\displaystyle}

\def\LL{{\mathcal L}}

\def\cl{\centerline}

\def\vs{\vspace*}

\def\ni{\noindent}

\def\Z{\mathbb{Z}}

\def\F{\mathbb{F}}
\def\QED{\hfill$\Box$}

\numberwithin{equation}{section}
\newtheorem{theo}{Theorem}[section]
\newtheorem{defi}[theo]{Definition}
\newtheorem{coro}[theo]{Corollary}
\newtheorem{lemm}[theo]{Lemma}

\usepackage[usenames]{color}

\begin{document}

\cl{{\large\bf Quantization of extended Schr\"{o}dinger-Virasoro Lie
algebra }\footnote{Supported by NSF grants 10825101 of China
\\ \indent $^{*}$Corresponding author:
lmyuan@mail.ustc.edu.cn}}\vs{6pt}

\cl{ Lamei Yuan$^{\,*}$, Liji Zhou$^{\,\dag}$}
 \cl{\small $^{*}$Department of Mathematics,
University of Science \!and \!Technology \!of \!China, Hefei 230026,
China} \cl{\small $^{\dag}$School of Mathematics and Computer
Science, Guizhou Normal University, Guizhou 550001, China}

 \cl{\small E-mail: lmyuan@mail.ustc.edu.cn, zhou\_li\_ji@126.com}\vs{6pt}

{\small\parskip .005 truein \baselineskip 3pt \lineskip 3pt

\noindent{{\bf Abstract.}  In present paper, we quantize the
extended Schr\"{o}dinger-Virasoro Lie algebra
 in characteristic zero with its Lie bialgebra structures classified by Yuan-Wu-Xu, and get a new Hopf algebra.
\vs{5pt}

\noindent{\bf Key words:} Lie bialgebras, Quantizaton, Extended
Schr\"{o}dinger-Virasoro Lie algebra.}

\parskip .001 truein\baselineskip 9pt \lineskip 9pt

\vs{18pt}

\cl{\bf\S1. \
Introduction}\setcounter{section}{1}\setcounter{equation}{0}

\vs{6pt}

The Schr\"{o}dinger-Virasoro Lie algebras \cite{H1} were introduced
in the context of non-equilibrium statistical physics during the
process of investigating  free Schr\"{o}dinger equations. They are
closely related to Schr\"{o}dinger algebra and Virasoro algebra.
 Later J.
Unterberger further constructed a new class of infinite-dimensional
Lie algebras called extended Schr\"{o}dinger-Virasoro Lie algebra
$\LL$ (see \rm {Definition 1.5} in [\ref{U}]), which can be viewed
as an extension of the original Schr\"{o}dinger-Virasoro Lie algebra
by a conformal current of weight one and is generated by $\{L_n,M_n,
N_n, Y_p\,|\,n \in \Z, \ p\in \Z+1/2\}$ with the following Lie
brackets:
\begin{eqnarray}\label{LB}\begin{array}{lll}
&&[L_m,L_{n}]=(n-m)L_{n+m},\ \ \ \ \ [L_m,N_n]=nN_{m+n},\ \ \ \ \,
[L_m,M_n\,]=nM_{n+m},\
\\[4pt]&&[L_n,\,Y_p\,]=(p-n/2)Y_{p+n},\ \ \ \ \, [N_m,Y_p\,]=Y_{m+p},\ \ \ \ \ \ \ \,[N_m,M_n]=2M_{m+n},\ \ \\ [4pt]&&
[M_n,Y_p\,]=[N_m,N_n]=0,\ \ \ \ \ \ [M_m,M_n]=0, \ \ \ \ \ \ \ \ \ \
\ \,[\,Y_p,Y_{q}\,\,]=(q-p)M_{p+q}.\end{array}
\end{eqnarray}
Note that $\LL$ is centerless and finitely generated by
$\{L_{-2},L_{-1},L_1,L_2,N_1,Y_{1/2}\}$. The derivations, central
extensions and automorphisms of $\LL$ have been studied in
[\ref{GJP}]. Recently, Yuan-Wu-Xu in \cite{YWX} classified all Lie
bialgebra structures on this kind of Lie algebras, which turned out
to be triangular coboundary (for the definition, see p.28,
\cite{ES}). This is different from that in Schr\"{o}dinger-Virasoro
algebra case \cite{HS}. \par

   As is known to all, constructing quantization of Lie bialgebras is an important
approach to producing new quantum groups. So this work has been
paying more attention in both mathematics and physics, and has been
investigated in a series of papers (see e.g.
[\ref{EK1}--\ref{EK2}]).

In present paper, we use the general quantization method by a
Drinfeld twist (cf.~[\ref{D1}]) to quantize explicitly the newly
determined triangular Lie bialgebra structures on the extended
Schr\"{o}dinger-Virasoro Lie algebra $\LL$ in characteristic zero
[\ref{YWX}]. Actually, this process completely depends on the
construction of Drinfeld twists, which, up to integral scalars, are
controlled by the classical Yang-Baxter $r$-matrix.  Our result
extends the class of examples of noncommutative and noncocommutative
Hopf algebras.
\par

 The main result of this paper is the following:
\begin{theo}\label{theo1}\vskip-3pt
With the choice of two distinguished elements $h=N_{0}$ and
$e=Y_{p}$ $(p\in\Z+1/2)$ such that $[h,e]=e$ in  $\mathcal {L}$,
there exists a structure of noncommutative and noncocommutative Hopf
algebra on $U(\mathcal {L})[[t]]$ denoted by $(U(\mathcal {L})[[t]],
m,\iota,\Delta, S,\epsilon)$ which leaves the product and counit of
$U(\mathcal {L})[[t]]$ undeformed but with a comultiplication and
antipode defined by:
\begin{eqnarray*}
&&\Delta{(L_n)}=1\otimes L_n+ L_n\otimes 1+(p-n/2)\big(h\otimes
(1-et)^{-1}Y_{n+p}t-n/2h^{(2)}\otimes (1-et)^{-2}M_{2p+n}t^2\big),\\[-6pt]
&&\Delta({N_n})=1\otimes N_n+ N_n\otimes 1+h\otimes
(1-et)^{-1}Y_{n+p}t-n/2h^{(2)}\otimes (1-et)^{-2}M_{2p+n}t^2,\\[-6pt]
&&{\Delta(M_n)}=1\otimes(1-et)^{2}\cdot(M_n\otimes 1)+1\otimes M_n
=M_n\otimes(1-et)^{2}+1\otimes M_n,\\[-6pt]
&&{\Delta(Y_q)}
=Y_q\otimes(1-et)+1\otimes Y_q+(p-q)h\otimes (1-et)^{-1}M_{p+q}t,\\[-6pt]
&&S(L_n)=-L_n+(p+n/2)t h(Y_{p+n}+n/2h^{(1)}_{-1}M_{2p+n}t),\\[-6pt]
&&S(Y_q)=-(1-et)^{-1}\big(Y_q+(p-q)h^{(1)}_{-1}M_{p+q}t\big),\\[-6pt]
&&S(N_n)=-N_n+t h(Y_{p+n}+n/2h^{(1)}_{-1}M_{2p+n}t),\\[-6pt]
&& S(M_n)=-(1-et)^{-2}\cdot M_n\,.
\end{eqnarray*}
\end{theo}

\vs{6pt}

\cl{\bf\S2. \ Some preliminary results}\setcounter{section}{2}
\setcounter{theo}{0}\setcounter{equation}{0} \vs{6pt}

 Throughout this
work, we denote by $\Z_+$ the set of all nonnegative integers,
$\mathbb{F}$ a field of characteristic zero and $\Z^*$ (resp.
$\F^*$) the set of all nonzero elements of $\Z$ (resp. $\F$).\par
 Let $(U(\mathcal {L}),m,\iota,\Delta_{0},S_{0},\epsilon)$ be the
standard Hopf algebra structure on $U(\mathcal {L})$, i.e.,
\begin{eqnarray}\label{2}
\Delta_{0}(X)=X\otimes 1+1\otimes X,& S_{0}(X)=-X,
&\epsilon(X)=0\mbox{ \ for }X\in{\mathcal{L}},
\end{eqnarray}
where $\Delta_{0}$ is a comultiplication, $\epsilon$ is a counit and
$S_0$ is an antipode. In particular,
$$\Delta_0(1)=1\otimes 1,\ \ \varepsilon(1)=S_0(1)=1.$$
\begin{defi}\label{def2}
Let $(H,m,\iota,\Delta_{0},S_{0},\epsilon)$ be a Hopf algebra over a
commutative ring $R$. A Drinfeld twist $\mathcal {F}$ on $H$ is an
invertible element of $H\otimes H$ such that
$$(\mathcal {F}\otimes 1)(\Delta_{0}\otimes {\rm Id})(\mathcal {F})=(1\otimes \mathcal {F})({\rm Id}\otimes\Delta_{0})(\mathcal
{F}), \ \ \ (\epsilon\otimes {\rm Id})(\mathcal {F})=1\otimes
1=({\rm Id}\otimes\epsilon)(\mathcal {F}).$$
\end{defi}\par
The following result is well known (see [\ref{D1}, \ref{ES}], etc.)

\begin{lemm}\label{Legr}\rm
Let $(H,m,\iota,\Delta_{0},S_{0},\epsilon)$ be a Hopf algebra over a
commutative ring and  $\mathcal {F}$ be a Drinfeld twist on $H$,
then $w=m({\rm Id}\otimes S_{0})(\mathcal {F})$ is invertible in $H$
with $w^{-1}=m(S_{0}\otimes {\rm Id})(\mathcal {F}^{-1})$. Moreover,
define $\Delta$ :$H\rightarrow H\otimes H$ and $S$ :$H\rightarrow H$
by
$$\begin{array}{llll}
\Delta(x)=\mathcal {F}\Delta_{0}(x)\mathcal {F}^{-1},&
S=wS_{0}(x)w^{-1},  \mbox{ \ for all \ }x\in H.
\!\!\!\!\!\!\!\!\!\!\!\!\!\!\!\!\!\!\!\!\!\!\!\!\end{array}$$ Then
 $(H,m,\iota,\Delta,S,\epsilon)$ is a new Hopf algebra, which is said to be the
 twisting of $H$ by the Drinfeld twist $\mathcal {F}$.
\end{lemm}\vs{-6pt}

For any element $x$ of a unital $R$-algebra ($R$ is a ring) and
$a\in R$, we set (see, e.g., [\ref {GZ}])
\begin{eqnarray*}
&&x^{(n)}_{a}:=(x+a)(x+a+1)\cdots(x+a+n-1),\\
&&x^{[n]}_{a}:=(x+a)(x+a-1)\cdots(x+a-n+1),
\end{eqnarray*}
and $x^{(n)}:=x^{(n)}_{0}$,  $x^{[n]}:=x^{[n]}_{0}$.

\begin{lemm}\label{Legr3}{\rm (see [\ref{GZ}, \ref{G}])} For any element $x$ of a
unital $\mathbb{F}$-algebra, $a,b\in \mathbb{F}$, and  $r,s,t\in
\mathbb{Z}$, one has
\begin{eqnarray}\label{fir-e}
&&x^{(s+t)}_{a}=x^{(s)}_{a}x^{(t)}_{a+s},\ \ \ \ \  x^{[s+t]}_{a}=
x^{[s]}_{a}x^{[t]}_{a-s},\ \ \ \ \
 x^{[s]}_{a}= x^{(s)}_{a-s+1},\label{fir-e}\\
&&\mbox{$\sum\limits_{s+t=r}$}\frac{(-1)^{t}}{s!t!}x^{[s]}_{a}x^{(t)}_{b}={a-b\choose
r}=
\frac{(a-b)\cdots(a-b-r+1)}{r!},\label{fir-e1}\\
&&\mbox{$\sum\limits_
{s+t=r}$}{\displaystyle\frac{(-1)^{t}}{s!t!}}x^{[s]}_{a}x^{[t]}_{b-s}={\displaystyle{a-b+r-1\choose
r}}=\frac{(a-b)\cdots(a-b+r-1)}{r!}.\label{fir-e2}
\end{eqnarray}
\end{lemm}\par
The following popular result will be frequently used in the third
part of this paper.
\begin{lemm}\rm \label{Legr4}(see e.g., \cite[Proposition 1.3(4)]{SF}) For any elements $x, y$ of an
associative algebra $A$, and $m\in\mathbb{Z}_{+}$, one has
\begin{eqnarray}
xy^{m}=\mbox{$\sum\limits_{k=0}^{m}$}(-1)^k{m\choose k}y^{m-k}({\rm
ad\,}y)^k(x).
\end{eqnarray}
\end{lemm}

\vs{6pt} \cl{\bf\S3. \ Proof of the main
results}\setcounter{section}{3}
\setcounter{theo}{0}\setcounter{equation}{0} \vs{6pt}

To describe a quantization of $U(\mathcal {L})$, we need to
construct explicitly a Drinfeld twist according to Lemma \ref{Legr}.
Set $h=N_0$ and $e=Y_p$ for a  fixed $p\in \mathbb{Z}+1/2$. The
following calculations are necessary for constructing the Drinfeld
twist element in the sequel.
\begin{lemm}\rm \label{lemm1}
For $a\in\mathbb{F}$, $i\in\mathbb{Z}_{+}$, $n\in\mathbb{Z}$ and
$q\in\mathbb{Z}+1/2$, one has
\begin{eqnarray*}\begin{array}{lll}
&&L_nh^{(i)}_a=h^{(i)}_{a}L_n,\ N_nh^{(i)}_a=h^{(i)}_aN_n,\
M_nh^{(i)}_a=h^{(i)}_{a-2}M_n,\ Y_qh^{(i)}_a=h^{(i)}_{a-1}Y_q,\
L_nh^{[i]}_a=h^{[i]}_aL_n,\\[6pt]
&&N_nh^{[i]}_a=h^{[i]}_aN_n,\ M_nh^{[i]}_{a}=h^{[i]}_{a-2}M_n,\
Y_qh_a^{[i]}=h^{[i]}_{a-1}Y_q,\ e^nh^{(i)}_{a}=h^{(i)}_{a-n}e^n,\
e^nh^{[i]}_{a}=h^{[i]}_{a-n}e^n.\end{array}
\end{eqnarray*}
\end{lemm}

\noindent{\it Proof.~}~We only prove the first equation in the Lemma
(the others can be obtained similarly). Since $L_nh=hL_n$, there is
nothing to prove for $i=1$. For the induction step, suppose that it
holds for $i$, then one has
$$L_nh^{(i+1)}_a=L_nh^{(i)}_a(h+a+i)=h^{(i)}_aL_n(h+a+i)=h^{(i)}_{a}(h+a+i)L_n=h^{(i+1)}_{a}L_n.
\eqno\Box$$\par For $a\in\mathbb{F}$, set
\begin{eqnarray}\label{fom1}
&\mathcal
{F}_a=\sum\limits^{\infty}_{r=0}\frac{(-1)^r}{r!}h^{[r]}_a\otimes
e^r t^r,& F_a=
\mbox{$\sum\limits^{\infty}_{r=0}$}\frac{1}{r!}h^{(r)}_a\otimes e^r
t^r,
\\
&u_a=m\cdot (S_0\otimes{ \rm Id})(F_a), &v_a=m\cdot({\rm Id}\otimes
S_0)(\mathcal {F}_a).\nonumber
\end{eqnarray}
Write $\mathcal {F}=\mathcal {F}_0$, $F=F_0$, $u=u_0$, $v=v_0$. Since $S_0(h^{(r)}_a)=(-1)^rh^{[r]}_{-a}$ and $S_0(e^r)=(-1)^r e^r$, we have
\begin{eqnarray}\label{fom2}
u_a=\mbox{$\sum\limits^{\infty}_{r=0}$}\frac{(-1)^r}{r!}h^{[r]}_{-a}e^r
t^r,&&v_a= \mbox{$\sum\limits^{\infty}_{r=0}$}\frac{1}{r!}h^{[r]}_a
e^r t^r.
\end{eqnarray}

\begin{lemm}\rm \label{lemm2}
For $a,b\in\mathbb{F}$, one has
\begin{eqnarray*}
\mathcal {F}_aF_b=1\otimes(1-et)^{a-b},&& v_au_b=(1-et)^{-(a+b)}.
\end{eqnarray*}
\end{lemm}
\noindent{\it Proof.}~~By  (\ref{fir-e1}) and (\ref{fom1}),
we have
\begin{eqnarray*}
\mathcal {F}_aF_b&=&
\mbox{$\sum\limits^{\infty}_{r,s=0}$}\frac{(-1)^r}{r!s!}h^{[r]}_ah^{(s)}_b\otimes
e^r e^s t^r t^s=\mbox{$\sum\limits^{\infty}_{m=0}$}(-1)^m
\big(\mbox{$\sum\limits_{r+s=m}$}\frac{(-1)^s}{r!s!}h^{[r]}_{a}h^{(s)}_b\big)\otimes e^m t^m\\
&=&\mbox{$\sum\limits^{\infty}_{m=0}$}(-1)^m{a-b \choose m} \otimes
e^m t^m=1\otimes(1-et)^{a-b}.
\end{eqnarray*}
It follows from (\ref{fir-e2}), (\ref{fom2}) and Lemma \ref{lemm1}
that
\\[4pt]
\hspace*{60pt}$v_au_b=\sum\limits^{\infty}_{r,s=0}{\displaystyle\frac{(-1)^s}{r!s!}}h^{[r]}_a
e^r h^{[s]}_{-b} e^s
t^{r+s}=\sum\limits^{\infty}_{m=0}\sum\limits_{r+s=m}{\displaystyle\frac{(-1)^s}{r!s!}}
h^{[r]}_{a}h^{[s]}_{-b-r}e^m t^m$
\\[4pt]
\hspace*{60pt}$\phantom{v_au_b}=\sum\limits^{\infty}_{m=0}{\displaystyle{a+b+m-1\choose
m}}e^m t^m=(1-et)^{-(a+b)}.$\QED
\begin{coro}\rm \label{coro}  For $a\in\mathbb{F}$, the elements $F_a$ and $u_a$
are invertible with $F^{-1}_a=\mathcal {F}_a$, $u^{-1}_a=v_{-a}$. In
particular, $F^{-1}=\mathcal{F}$, $u^{-1}=v$.
\end{coro}
\begin{lemm}\rm \label{lemm3}
For any $a\in\mathbb{F}$ and $r\in\mathbb{Z}_+$, one has $
\Delta_0(h^{[r]})=\mbox{$\sum\limits^r_{i=0}$}{r\choose
i}h^{[i]}_{-a}\otimes h^{[r-i]}_{a}. $
 In particular, one has $\Delta_0(h^{[r]})=\mbox{$\sum\limits^r_{i=0}$}{r\choose i}h^{[i]}\otimes
 h^{[r-i]}$.
\end{lemm}
\noindent{\it Proof.}\ \ It can be proved by induction on $r$. For
$r=1$, both sides in the formula are equal to $1\otimes h+h\otimes
1$. Assume that it is true for $r>1$, then we have
\begin{eqnarray*}
\D_0(h^{[r+1]})&=&\D_0(h^{[r]}(h-r))=\D_0(h^{[r]})\big(\D_0(h)-\D_0(r)\big)\\
&=&\Big(\mbox{$\sum\limits^r_{i=0}$}{r\choose i}h^{[i]}_{-a}\otimes
h^{[r-i]}_a\Big)\big((h-r)\otimes 1+1\otimes (h-r)+r(1\otimes 1)\big)\\
&=&\Big(\mbox{$\sum\limits^{r-1}_{i=1}$}{r\choose
i}h^{[i]}_{-a}\otimes h^{[r-i]}_a\Big)\big((h-r)\otimes 1+1\otimes
(h-r)\big)+r\mbox{$\sum\limits^r_{i=0}$}{r\choose
i}h^{[i]}_{-a}\otimes h^{[r-i]}_a\\
&\ &+h^{[r+1]}_{-a}\otimes 1+h^{[r]}_{-a}\otimes
a+h^{[r]}_{-a}\otimes(h-r)+(h-r)\otimes h^{[r]}_a+1\otimes
h^{[r+1]}_a-a\otimes h^{[r]}_a\\
&=&1\otimes h^{[r+1]}_a+h^{[r+1]}_{-a}\otimes
1+r\mbox{$\sum\limits^{r-1}_{i=1}$}{r\choose i}h^{[i]}_{-a}\otimes
h^{[r-i]}_a+h^{[r]}_{-a}\otimes (h+a)+(h-a)\otimes h^{[r]}_a\\
&\ &+\mbox{$\sum\limits_{i=1}^{r-1}$}{r\choose
i}h^{[i+1]}_{-a}\otimes
h^{[r-i]}_a+\mbox{$\sum\limits_{i=1}^{r-1}$}(-r+a+i){r\choose
i}h^{[i]}_{-a}\otimes
h^{[r-i]}_a\\
&\ &+\mbox{$\sum\limits_{i=1}^{r-1}$}{r\choose i}h^{[i]}_{-a}\otimes
h^{[r-i+1]}_a+\mbox{$\sum\limits_{i=1}^{r-1}$}(-a-i){r\choose
i}h^{[i]}_{-a}\otimes h^{[r-i]}_a\\
&=&1\otimes h^{[r+1]}_{a}+h^{[r+1]}_{-a}\otimes
1+\mbox{$\sum\limits_{i=1}^{r}$}\bigg[{r\choose i-1}+{r\choose
i}\bigg]h^{[i]}_{-a}\otimes h^{[r-i+1]}_a\\
&=&\mbox{$\sum\limits_{i=0}^{r+1}$}{r+1 \choose
i}h^{[i]}_{-a}\otimes h^{[r+1-i]}_a.
\end{eqnarray*}
Therefore, the formula holds by induction. \QED
\begin{lemm}\rm \label{lemm5}
The element $\mathcal
{F}=\mbox{$\sum\limits^{\infty}_{r=0}$}\frac{(-1)^r}{r!}h^{[r]}\otimes
e^rt^r$ is a Drinfeld twist on $U(\mathcal {L})[[t]]$ .
\end{lemm}
\noindent{\it Proof.}\ \ It can be proved directly by the similar
methods as those presented in the proof of \cite[Proposition
2.5]{HW}.\QED\vskip4pt\par Now we can perform the process of
twisting the standard Hopf structure $(U(\mathcal
{L}),m,\iota,\Delta_0, S_0,\epsilon)$ defined in (\ref{2}) by the
Drinfeld twist $\mathcal {F}$ constructed above. The following
lemmas are very useful to our main results.
\begin{lemm}\rm \label{lemm6}   For $a\in\mathbb{F}$, $n\in\mathbb{Z}$, $q\in\mathbb{Z}+1/2$,
one has
\begin{eqnarray*}\begin{array}{lll}
(L_n\otimes 1)F_a=F_{a}(L_n\otimes 1),&& (N_n\otimes
1)F_a=F_a(N_n\otimes 1),\\[6pt]
(M_n\otimes 1)F_a=F_{a-2}(M_n\otimes1), &&(Y_q\otimes
1)F_a=F_{a-1}(Y_q\otimes 1).\end{array}
\end{eqnarray*}
\end{lemm}
\noindent{\it Proof.}\ \ It follows directly from (\ref{fom1}) and
Lemma \ref{lemm1}.\QED
\begin{lemm}\rm \label{lemm7}
For $n\in\Z,\ \ q\in\Z+1/2$ and $r\in\mathbb{Z}_+$, one has
\begin{eqnarray}
&&L_ne^r=e^rL_n+(p-n/2)(re^{r-1}Y_{p+n}-nr(r-1)/2e^{r-2}M_{2p+n}),\label{fom4}\\
&&N_ne^r=e^rN_n+re^{r-1}Y_{p+n}-nr(r-1)/2e^{r-2}M_{2p+n},\label{fom5}\\
 &&Y_qe^r=e^rY_q+r(p-q)e^{r-1}M_{p+q},\ \ \ \
 M_ne^r=e^rM_n\label{fom6}.\ \ \ \ \ \ \ \
\end{eqnarray}
\end{lemm}
\noindent{\it Proof.}\ \ By Lemma \ref{Legr4} and (\ref{LB}), we
have
\begin{eqnarray*}
L_ne^r&=&\mbox{$\sum\limits^r_{i=0}$}(-1)^i{r\choose i}e^{r-i}({\rm
ad\,} e)^{i}(L_n)\\
&=&e^rL_n+(p-n/2)(re^{r-1}Y_{p+n}-nr(r-1)/2e^{r-2}M_{2p+n}).
\end{eqnarray*}
Similarly, one can get (\ref{fom5}) and (\ref{fom6}). \QED
\begin{lemm}\rm \label{lemm8}  For $a\in\mathbb{F}$, $n\in\mathbb{Z}$, and $q\in\mathbb{Z}+1/2$, we have
\begin{eqnarray}
(1\otimes L_n)F_a =F_a(1\otimes
L_n)+(p-n/2)\big(F_{a+1}(h_a^{(1)}\otimes
Y_{n+p})t-n/2F_{a+2}(h_a^{(2)}\otimes M_{2p+n})t^2\big),\label{fom7}\\
(1\otimes N_n)F_a =F_a(1\otimes N_n)+F_{a+1}(h_a^{(1)}\otimes
Y_{n+p})t-n/2F_{a+2}(h_a^{(2)}\otimes M_{2p+n})t^2,\ \ \ \ \ \ \ \ \ \ \ \ \ \ \label{fom8}\\
(1\otimes Y_q)F_a=F_a(1\otimes Y_q)+(p-q)F_{a+1}(h^{(1)}_a\otimes
M_{p+q})t,\ \ \ \ \ \ (1\otimes M_n)F_a=F_a(1\otimes
M_n).\label{fom9}\
\end{eqnarray}
\end{lemm}

\noindent{\it Proof.}\ \  By (\ref{fir-e}), (\ref{fom1}) and
(\ref{fom4}), one has
\\[4pt]\hspace*{4ex}$\dis
 (1\otimes L_n)F_a=\mbox{$\sum\limits^{\infty}_{r=0}$}\frac{1}{r!}h^{(r)}_a\otimes L_n e^r
 t^r$\\[4pt]\hspace*{4ex}$\dis\phantom{ (1\otimes L_n)F_a}
 =\mbox{$\sum\limits^{\infty}_{r=0}$}\frac{1}{r!}h_a^{(r)}\otimes\big(e^rL_n+(p-n/2)(re^{r-1}Y_{p+n}-\frac{nr(r-1)}{2}e^{r-2}M_{2p+n}\big)t^r
$\\[4pt]\hspace*{4ex}$\dis\phantom{ (1\otimes L_n)F_a}
 =F_a(1\otimes
 L_n)+(p-n/2)\big(\mbox{$\sum\limits^{\infty}_{r=1}$}\frac{1}{(r-1)!}
h^{(r)}_{a}\otimes e^{r-1}Y_{n+p}-\frac{n}{2}\mbox{$\sum\limits^{\infty}_{r=2}$}\frac{1}{(r-2)!}h^{(r)}_a\otimes e^{r-2}M_{2p+n}\big) t^{r}$\\[4pt]\hspace*{4ex}$\dis\phantom{ (1\otimes L_n)F_a}
 =F_a(1\otimes
 L_n)+(p-n/2)\big(\mbox{$\sum\limits^{\infty}_{r=0}$}\frac{1}{r!}
h^{(r+1)}_{a}\otimes
e^{r}Y_{n+p}t^{r+1}-\frac{n}{2}\mbox{$\sum\limits^{\infty}_{r=0}$}\frac{1}{r!}h^{(r+2)}_a\otimes
e^{r}M_{2p+n}t^{r+2}\big)
$\\[4pt]\hspace*{4ex}$\dis\phantom{ (1\otimes L_n)F_a}
=F_a(1\otimes
 L_n)+(p-n/2)\big(\mbox{$\sum\limits^{\infty}_{r=0}$}\frac{1}{r!}
h^{(r)}_{a+1}h_a^{(1)}\otimes
e^{r}Y_{n+p}t^{r+1}-\frac{n}{2}\mbox{$\sum\limits^{\infty}_{r=0}$}\frac{1}{r!}h^{(r)}_{a+2}h_a^{(2)}\otimes
e^{r}M_{2p+n}t^{r+2}\big)$\\[4pt]\hspace*{4ex}$\dis\phantom{ (1\otimes L_n)F_a}
 =F_a(1\otimes L_n)+(p-n/2)\big(F_{a+1}\big(h^{(1)}_a\otimes
 Y_{n+p}\big)t-n/2F_{a+2}\big(h^{(2)}_{a}\otimes
 M_{2p+n}\big)t^2\big)$.\\
This proves equation (\ref{fom7}). Similarly, (\ref{fom8}) and
(\ref{fom9}) follow from (\ref{fom5}) and (\ref{fom6}),
respectively.  \QED
\begin{lemm}\rm \label{lemm9}
 For $a\in\mathbb{F}$, $n\in\mathbb{Z}$, $q\in\frac{1}{2}+\mathbb{Z}$,
 one has
 \begin{eqnarray}
 L_n u_a=u_aL_n-(p-n/2)tu_ah^{(1)}_{-a}\big(Y_{p+n}+tn/2h^{(1)}_{-a-1}M_{2p+n}\big),\label{fom10}\\
 N_n u_a=u_aN_n-tu_ah^{(1)}_{-a}\big(Y_{p+n}+tn/2h^{(1)}_{-a-1}M_{2p+n}\big),\ \ \ \ \ \ \ \ \ \ \ \ \label{fom11}\\
 Y_qu_a=u_{a+1}Y_q+(q-p)tu_{a+1}h_{-a-1}^{(1)}M_{p+q},\ \
 M_nu_a=u_{a+2}M_n.\label{fom12}
 \end{eqnarray}
\end{lemm}
 \noindent{\it Proof.}\ \
From equation (\ref{fir-e}), (\ref{fom2}),  (\ref{fom4})  and Lemma
\ref{lemm1}, one has
\\[4pt]\hspace*{4ex}$\dis
 L_nu_a=\mbox{$\sum\limits^{\infty}_{r=0}$}\frac{(-1)^r}{r!}L_nh^{[r]}_{-a}e^rt^r=\mbox{$\sum\limits^{\infty}_{r=0}$}\frac{(-1)^r}{r!}h^{[r]}_{-a}L_ne^rt^r$\\
 [4pt]\hspace*{4ex}$\dis\phantom{L_nu_a}=\mbox{$\sum\limits^{\infty}_{r=0}$}\frac{(-1)^r}{r!}h^{[r]}_{-a}\big(e^rL_n+(p-n/2)(re^{r-1}Y_{p+n}-nr(r-1)/2e^{r-2}M_{2p+n})\big)t^r$\\
[4pt]\hspace*{4ex}$\dis\phantom{L_nu_a}=u_aL_n+(p-n/2)\big(\mbox{$\sum\limits^{\infty}_{r=1}$}\frac{(-1)^r}{(r-1)!}h^{[r]}_{-a}e^{r-1}t^{r}Y_{p+n}-n/2\mbox{$\sum\limits^{\infty}_{r=2}$}\frac{(-1)^r}{(r-2)!}h^{[r]}_{-a}e^{r-2}t^{r}M_{2p+n}\big)$\\
[4pt]\hspace*{4ex}$\dis\phantom{L_nu_a}
=u_aL_n-(p-n/2)\big(\mbox{$\sum\limits^{\infty}_{r=0}$}\frac{(-1)^r}{r!}h^{[r+1]}_{-a}e^{r}t^{r+1}Y_{p+n}+n/2\mbox{$\sum\limits^{\infty}_{r=0}$}\frac{(-1)^r}{r!}h^{[r+2]}_{-a}e^{r}t^{r+2}M_{2p+n}\big)$\\
[4pt]\hspace*{4ex}$\dis\phantom{L_nu_a}
=u_aL_n-(p-n/2)\big(\mbox{$\sum\limits^{\infty}_{r=0}$}\frac{(-1)^r}{r!}h^{[r]}_{-a}h^{[1]}_{-a-r}e^{r}t^{r+1}Y_{p+n}+n/2\mbox{$\sum\limits^{\infty}_{r=0}$}\frac{(-1)^r}{r!}h^{[r]}_{-a}h^{[2]}_{-a-r}e^{r}t^{r+2}M_{2p+n}\big)$\\
[4pt]\hspace*{4ex}$\dis\phantom{L_nu_a}
=u_aL_n-(p-n/2)\big(\mbox{$\sum\limits^{\infty}_{r=0}$}\frac{(-1)^r}{r!}h^{[r]}_{-a}e^{r}h^{[1]}_{-a}t^{r+1}Y_{p+n}+n/2\mbox{$\sum\limits^{\infty}_{r=0}$}\frac{(-1)^r}{r!}h^{[r]}_{-a}e^{r}h^{[2]}_{-a}t^{r+2}M_{2p+n}\big)$\\
[4pt]\hspace*{4ex}$\dis\phantom{L_nu_a}
 =u_aL_n-(p-n/2)tu_ah^{(1)}_{-a}\big(Y_{p+n}+tn/2h^{(1)}_{-a-1}M_{2p+n}\big).$\\
 Hence, (\ref{fom10}) holds. Similarly, one can get (\ref{fom11}) and
 (\ref{fom12}) by Lemma \ref{lemm1} and Lemma \ref{lemm7}.
 \QED

Now we have enough in hand to proof our main Theorem in this paper.

\ni{\it Proof of Theorem \ref{theo1}} \ \ \rm\  Thanks to Lemma
\ref{Legr},
Lemma \ref{lemm2}, Corollary \ref{coro}, Lemma \ref{lemm6} and Lemma \ref{lemm8}, we have\\
[4pt]\hspace*{4ex}$\dis
 \Delta(L_n)=\mathcal
{F}\cdot \Delta_0(L_n)\cdot \mathcal{F}^{-1}=\mathcal {F}\cdot (L_n\otimes
1)\cdot F+\mathcal {F}\cdot (1\otimes L_n)\cdot F$\\
[4pt]\hspace*{4ex}$\dis\phantom{\Delta(L_n)} =\mathcal {F}\cdot
F\cdot L_n\otimes 1+\mathcal{F}\cdot(F\cdot 1\otimes
L_n+(p-n/2)(F_1\cdot h^{(1)}\otimes
Y_{n+p}t-n/2F_2\cdot h^{(2)}\otimes M_{2p+n}t^2)\big)$\\
[4pt]\hspace*{4ex}$\dis\phantom{\Delta(L_n)} =L_n\otimes 1+1\otimes
L_n+(p-n/2)\big(1\otimes(1-et)^{-1}\cdot h^{(1)}\otimes
Y_{n+p}t-n/2\otimes (1-et)^{-2}\cdot h^{(2)}\otimes M_{2p+n}t^2\big)$\\
[4pt]\hspace*{4ex}$\dis\phantom{\Delta(L_n)} =1\otimes L_n+
L_n\otimes 1+(p-n/2)\big(h\otimes
(1-et)^{-1}Y_{n+p}t-n/2h^{(2)}\otimes (1-et)^{-2}M_{2p+n}t^2\big).
$\\
[4pt]\hspace*{4ex}$\dis\Delta(N_n)=\mathcal {F}\cdot
\Delta_0(N_n)\cdot \mathcal{F}^{-1}=\mathcal {F}\cdot (N_n\otimes
1)\cdot F+\mathcal {F}\cdot (1\otimes N_n)\cdot F$\\
[4pt]\hspace*{4ex}$\dis\phantom{\Delta(N_n)} =\mathcal {F}\cdot
F\cdot N_n\otimes 1+\mathcal{F}\cdot(F\cdot 1\otimes L_n+F_1\cdot
h^{(1)}\otimes
Y_{n+p}t-n/2F_2\cdot h^{(2)}\otimes M_{2p+n}t^2\big)$\\
[4pt]\hspace*{4ex}$\dis\phantom{\Delta(N_n)} =1\otimes N_n+
N_n\otimes 1+h\otimes (1-et)^{-1}Y_{n+p}t-n/2h^{(2)}\otimes
(1-et)^{-2}M_{2p+n}t^2.
$\\[4pt]\hspace*{4ex}$\dis\Delta(M_n)=\mathcal{F}\cdot (M_n\otimes
1)\cdot F+\mathcal{F}\cdot (1\otimes M_n)\cdot F =\mathcal{F}\cdot
F_{-2}\cdot (M_n\otimes 1)+\mathcal{F}\cdot F(1\otimes M_n)$
\\[4pt]\hspace*{4ex}$\dis\phantom{\Delta(M_n)}
=1\otimes(1-et)^{2}\cdot(M_n\otimes 1)+1\otimes M_n
=M_n\otimes(1-et)^{2}+1\otimes M_n. $\\
[4pt]\hspace*{4ex}$\dis \Delta(Y_q)=\mathcal {F}\cdot
\Delta_0(Y_q)\cdot \mathcal{F}^{-1}=\mathcal{F}\cdot Y_q\otimes
1\cdot F+\mathcal{F}\cdot 1\otimes Y_q\cdot F$\\
[4pt]\hspace*{4ex}$\dis\phantom{\Delta(Y_q)}=\mathcal{F}\cdot
F_{-1}\cdot Y_q\otimes 1+\mathcal{F}\cdot(F\cdot1\otimes
Y_q+(p-q)F_1\cdot h\otimes
M_{p+q}t)$\\[4pt]\hspace*{4ex}$\dis\phantom{\Delta(Y_q)}
=1\otimes (1-et)\cdot Y_q\otimes 1+1\otimes Y_q+(p-q)\otimes
(1-et)^{-1}\cdot h\otimes M_{p+q}t$\\[4pt]\hspace*{4ex}$\dis\phantom{\Delta(Y_q)}
=Y_q\otimes(1-et)+1\otimes Y_q+(p-q)h\otimes (1-et)^{-1}M_{p+q}t.$\\
Again by Lemma \ref{Legr}, Corollary \ref{coro}, Lemma \ref{lemm2}
and Lemma \ref{lemm9}, we have
\begin{eqnarray*}
S(L_n)&=&u^{-1}S_0(L_n)u=-vL_nu\\[-6pt]
&=&-v\big(uL_n-(p-n/2)t u
h^{(1)}(Y_{p+n}+tn/2h^{(1)}_{-1}M_{2p+n})\big)\\[-6pt]
&=&-L_n+(p+n/2)t h(Y_{p+n}+n/2h^{(1)}_{-1}M_{2p+n}t).\\[-6pt]
S(N_n)&=&u^{-1}S_0(N_n)u=-vN_nu\\[-6pt]
&=&-v\big(uN_n-t u
h^{(1)}(Y_{p+n}+tn/2h^{(1)}_{-1}M_{2p+n})\\[-6pt]
&=&-N_n+t h(Y_{p+n}+n/2h^{(1)}_{-1}M_{2p+n}t).\\[-6pt]
S(M_n)&=&u^{-1}S_0(M_n)u=-v\cdot M_nu=-v\cdot
u_{2}M_n=-(1-et)^{-2}M_n.\\[-6pt]
S(Y_q)&=&u^{-1}S_0(Y_q)u=-vY_qu =-v\cdot\big(
u_{1}\cdot Y_q+(q-p)tu_1h^{(1)}_{-1}M_{p+q}\big)\\[-6pt]
&=&-(1-et)^{-1}\big(Y_q+(p-q)h^{(1)}_{-1}M_{p+q}t\big).\\[-6pt]
\end{eqnarray*}
So the proof is complete!\QED

\parskip=-1pt\baselineskip=3pt\lineskip=3pt

\end{document}